\documentclass[prb,twocolumn,showpacs,floatfix,,amsmath,amssymb]{revtex4}
\usepackage{amsfonts}
\usepackage{stmaryrd}
\usepackage{bbm}
\usepackage{mathrsfs}
\usepackage{tipa}
\usepackage{amssymb}
\usepackage{txfonts}
\usepackage{graphicx}
\usepackage{dcolumn}
\usepackage{epstopdf}
\usepackage[colorlinks,linkcolor=blue,urlcolor=blue,citecolor=blue]{hyperref}
\usepackage{multirow}

\begin{document}
\newcommand*{\cm}{cm$^{-1}$\,}
\newcommand*{\Tc}{T$_c$\,}

\title{Revealing a new charge density wave order in TbTe$_3$ by optical conductivity and ultrafast pump-probe experiments}
\author{R. Y. Chen, B. F. Hu, T. Dong, N. L. Wang}
\affiliation{Beijing National Laboratory for Condensed Matter
Physics, Institute of Physics, Chinese Academy of Sciences,
Beijing 100190, China}

\begin{abstract}
Rare-earth tri-tellurium RTe$_3$ is a typical quasi-two dimensional system which exhibits obvious charge density wave (CDW) orders. So far, RTe$_3$ with heavier R ions (Dy, Ho, Er and Tm) are believed to experience two CDW phase transitions, while the lighter ones only hold one. TbTe$_3$ is claimed to belong to the latter. However in this work we present evidences that TbTe$_3$ also possesses more than one CDW order. Aside from the one at 336 K, which was extensively studied and reported to be driven by imperfect Fermi surface nesting with a wave vector $q=(2/7 c^*)$, a new CDW energy gap (260 meV) develops at around 165 K, revealed by both infrared reflectivity spectroscopy and ultrafast pump-probe spectroscopy. More intriguingly, the origin of this energy gap is different from the second CDW order in the heavier R ions-based compounds RTe$_3$ (R=Dy, Ho, Er and Tm).
\end{abstract}

\pacs{78.30.Er, 78.40.Kc, 75.50.Cc}

\maketitle

\section{introduction}
Collective phenomena, such as density waves and superconductivity, are of fundamental importance in condensed matter physics. The coexistence and competition between them are among the key issues in a number of highly interested materials. Charge density wave is a symmetry broken state of metal induced by electron-phonon interactions. As the name implies, the charge density along with the lattice distortion are periodically modulated at its ground state. CDW is predominantly driven by nesting of Fermi surface (FS) which requires the presence of two patches of almost parallel FSs connected by a wave vector $\textbf{\emph{q}}_{CDW}$. A single particle energy gap would form in the nested region of the FSs. CDW instability usually appears in low dimensional materials, e.g. one-dimensional (1D) K$_0.3$MoO$_3$,\cite{PhysRevB.30.1971} 2D transition metal dichalcogenides,\cite{WU31031989} whose FSs are relatively simple and in favor of FS nesting.

RTe$_3$ is a typical quasi-2D material, with one layer of corrugated RTe slab and two sheets of square Te planes stacking along the \emph{b}-axis.\cite{doi:10.1021/ic50043a004} It is of orthorhombic structure, but the lattice parameter \emph{a} and \emph{c} only have tiny difference. Band structure calculation reveals that the FS is only associated with Te planes and the diamond shape of it is highly appropriate for nesting.\cite{PhysRevB.71.085114} CDW ordering was first discovered by transmission electron microscopy (TEM) in this system.\cite{PhysRevB.52.14516} Then a bunch of following experiments, such as X-ray diffraction (XRD)\cite{doi:10.1021/cm00046a049,PhysRevB.77.035114}, scanning tunneling microscopy (STM)\cite{PhysRevB.79.085422} and angle resolved photoemission spectroscopy (ARPES)\cite{PhysRevB.77.235104,PhysRevLett.81.886}, confirmed the conclusion and indicated that CDW transition commonly appears in this system at quite high temperature, driven by FS nesting with an incommensurate wave vector $\emph{\textbf{q}}=(2/7 \textbf{c}^*)$. Furthermore, by traversing R ions from La to Tm, the chemical pressure of RTe$_3$ increases due to the decreasing of R radii and the electronic structure alters correspondingly. As a consequence, the CDW transition temperature increases monotonically as the mass of R ions decreasing. For the lightest ones, the transition temperatures are expected to be even higher than their melting temperature.

Among the large family of RTe$_3$, the ones with heavier R ions (Dy, Ho, Er, Tm) are illustrated to experience two CDW transitions,\cite{PhysRevB.81.073102} while the lighter ones only undergo one.\cite{PhysRevB.77.035114} The second transitions occur at lower temperatures with a nesting wave vector whose magnitude is almost equal to the first one but orients at perpendicular direction, along \emph{$\textbf{a}^*$}-axis. Moreover, the transition temperature evolves in an opposite trend in contrast to the first one. Angle resolved photoemission spectroscopy (ARPES) reveals that the first CDW transition is of imperfect nesting, thus a second one is possible due to nesting of the remaining FS. Apart from that, the gapped area of FS caused by the first transition decreases as R radium decreasing. Therefore, the available area for the second one increases, which leads to the increasing of phase transition temperature.

As the nearest neighbor of DyTb$_3$, which experiences two CDW transitions, TbTe$_3$ is believed to hold only one near 335 K, illustrated by various techniques such as infrared reflectivity spectroscopy,\cite{PhysRevB.74.125115} XRD,\cite{PhysRevB.77.035114} and ultrafast pump-probe spectroscopy.\cite{PhysRevLett.101.246402} Recently, a report\cite{PhysRevB.87.155131} on TbTe$_3$ using high resolution XRD claimed that the compound experiences a second CDW transition at 41.0$\pm$0.4 K with a nesting wave vector along \emph{$\textbf{a}^*$}-axis, following the evolution trend of the other heavier RTe$_3$s, and the two CDW orders coexist with each other independently. In order to shed light on the underlying mechanism of coexist and competition between different CDW instabilities, it is essential to investigate the CDW transitions and evolution in TbTe$_3$ more scrupulously. Here, we use optical spectroscopy combined with ultrafast pump-probe spectroscopy, both of which are powerful techniques for detecting energy gaps in bulk materials, to probe the charge carrier dynamics. Unexpectedly, our measurements reveal the presence of a new CDW order near 165 K, besides the first one seen above the room temperature. This new CDW order does not follow the trend of second CDW order observed for the heavy rare-earth element based RTe$_3$ compounds in both transition temperature and energy scale of CDW gap. Therefore, it is different from the second CDW order reported by Banerjee et al.\cite{PhysRevB.87.155131}.

\section{results and discussion}
Plate-like single crystals with shining surfaces are grown by self-flux method. Figure 1 shows the in-plane dc resistivity as function of temperature T, which was measured using a standard four-probe method in a physical property measurement system. At high temperature the resistivity shows metallic behavior until a sudden upturn appears at $T_{C1}$, a signature of CDW phase transition.\cite{PhysRevB.77.035114} Blow $T_{C1}$ the resistivity approaches to a maximum around 300 K. Then metallic temperature dependence is observed once again at lower temperature and no anomalies ever exist down to 2 K, which implies that there seems to be only one phase transition. With the cooling and warming cycle, no hysteresis is observed.

\begin{figure}[tbp]
\includegraphics[width=7.5cm]{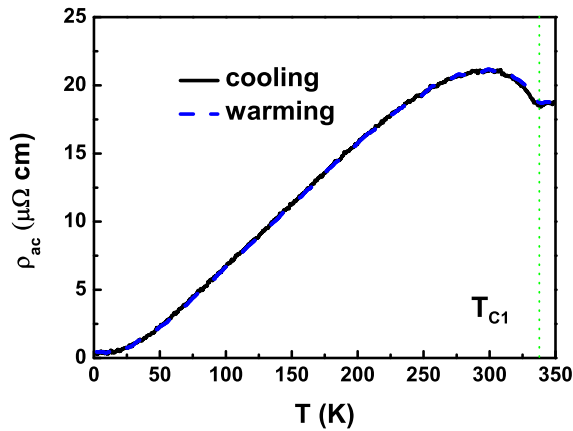}
\caption{(Color online) Temperature dependent resistivity of TbTe$_3$. Cooling and warming cycles are shown by solid and dashed lines respectively. $T_{C1}$ is marked by the green dotted line.}
\label{Fig:4}
\end{figure}

The optical reflectance measurement was performed on a combination of Bruker Vertex 80V and 113V spectrometers in the frequency range 40-25000 \cm. An \emph{in situ} gold and aluminum overcoating technique was used to get the reflectance R ($\omega$). The real part of conductivity $\sigma_1(\omega)$ is obtained by the Kramers-Kronig transformation of R($\omega$).

The main panel of Fig. 2 (a) shows the in-plane reflectance up to 5000 \cm at various temperatures, while the inset displays the expanded spectra up to 22000 \cm at two selected temperatures. Metallic behavior is clearly observed from both temperature and frequency dependent change, in agreement with the resistivity measurement. The reflectance at low frequency is very close to unit and increases upon cooling, revealing that TbTe$_3$ is a good metal. The most significant feature in spectra $R(\omega)$ is the substantial suppression in the mid-infrared region. At the highest temperature, $R(\omega)$ evolves rather smoothly below 5000 \cm and approaches to unit at zero frequency. Upon temperature cooling, we notice that a broad dip gradually emerges between 1500 \cm and 4500 \cm; it is quite weak at room temperature but gets more dramatic at lower temperatures. Additionally, the center of the dip shifts to higher frequency with temperature decreasing, around 3000 \cm at 10 K. The above mentioned characters are strong evidences for the formation of an energy gap. Apart from that, a shoulder like feature located roughly at 2000 \cm appears at 100 K and becomes more pronounced at 10K. This infrared absorption feature is indicative of another energy gap.

\begin{figure}[htbp]
\setlength{\belowcaptionskip}{0pt}
\includegraphics[width=7.5cm]{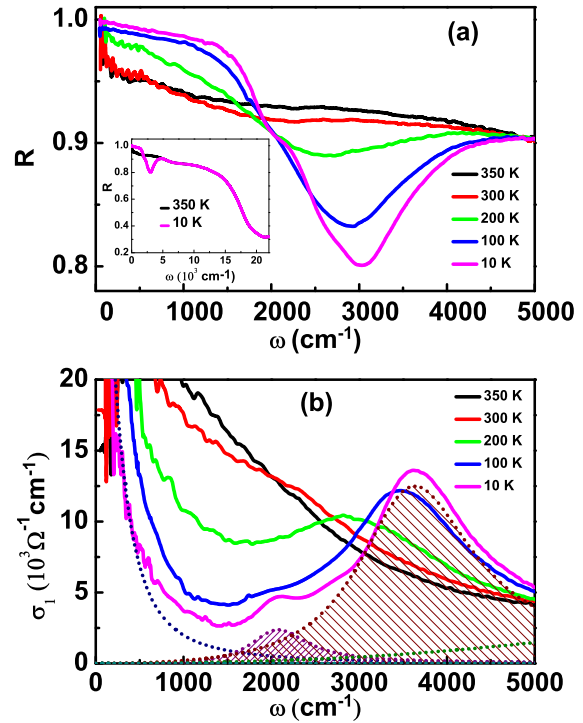}
\caption{(Color online) (a) Temperature dependent reflectivity of TbTe$_3$ below 5000 \cm. The inset shows $R(\omega)$ in an expanded range up to 22000 \cm at 10 K and 350 K. (b) Temperature dependent optical conductivity below 5000 \cm. The short dotted lines display the Drude-Lorentz fit of 10 K.}
\label{Fig:4}
\end{figure}

The conductivity spectra $\sigma_1(\omega)$ is plotted in Fig. 2 (b) with solid lines. Corresponding to the good metal property, the Drude component exists in the whole measurement temperature range. With temperature lowering, however, part of its spectral weight is removed to higher frequencies due to the opening of energy gap and leads to a broad peak around 3000 \cm. For a density wave phase transition, the spectral feature in frequency-dependent conductivity has been well established. The opening of an energy gap would cause a pronounced peak just above the energy gap in $\sigma_1(\omega)$, primarily due to the effect of the case-I coherent factor for density wave order. Consequently, we ascribe the strong suppression in $R(\omega)$ and the corresponding peak in $\sigma_1(\omega)$  to a CDW phase transition. Furthermore, the energy scale of the gap can be identified roughly at the peak position in $\sigma_1(\omega)$, which locates at 3600 \cm (450 meV). Along with the phase transition temperature, we can get the ratio $2\Delta_1/\kappa_B T_{C1}=15.8$, much higher than the weak coupling theory prediction.

\begin{table*}[bhtp]
\setlength\abovecaptionskip{0.5pt}
\caption{Fitting parameters of $\sigma_1(\omega)$ for 350 K and 10 K. $\omega_P$ is the plasma frequency and $\gamma_D=1/\tau$ is the scattering rate of free carriers. $\omega_j$, $\gamma_i=1/\tau_j$ and $S_j$ represent for the resonance frequency, the width and the square root of the oscillator strength of Lorentz terms, respectively. The unit of all the fitting parameters is ($\times$ 1000).  \label{1}}
\vspace{-1em}
\begin{center}
\renewcommand\arraystretch{1.5}
\begin{tabular}{p{1.2cm} p{1cm} p{1cm} p{1cm} p{1cm} p{1cm} p{1cm} p{1cm} p{1cm} p{1cm} p{1cm} p{1cm} p{1cm} p{1cm} p{0.6cm}}
\hline
\hline
 &$\omega_p$&$\gamma_D$&$\omega_1$&$\gamma_1$&$S_1$&$\omega_2$&$\gamma_2$&$S_2$&$\omega_3$&$\gamma_3$&$S_3$&$\omega_4$&$\gamma_4$&$S_4$\\
\hline
350 K&50&1.53&&&&&&&7.0&7.0&28&24&10&33\\
10 K&23&0.23&2.1&0.79&11&3.6&1.6&35&7.0&6.9&30&24&10&33\\
\hline
\hline
\end{tabular}\\
\end{center}
\end{table*}

Notably, for 10 K and 100 K an additional peak develops around 2100 \cm in associate with the shoulder-like feature in $R(\omega)$ spectra. As a consequence, there should be a second CDW gap opening in the temperature range between 200K and 100K. Unfortunately the infrared reflectivity spectroscopy is incapable of determining the phase transition temperature. Nevertheless the energy scale of this gap can be obtained as $2\Delta_2$ = 2100 \cm (260 meV). Combined with the phase transition temperature identified by our pump-probe measurement to be 165 K, the ratio of $2\Delta_2/\kappa_BT_{C2}=18.3$ is even higher than the first transition.

In order to get more information quantitatively, we use Drude-Lorentz model to decompose the conductivity $\sigma_1(\omega)$ into different components:

\begin{equation}
\epsilon(\omega)= \epsilon_{\infty}-\frac{\omega_{p}^2}{\omega^2+i\omega/\tau}+ \sum_{j}{\frac{S_j^2 }{\omega_j^2-\omega^2-i\omega/\tau_j}}. 
\end{equation}
Here, $\epsilon_{\infty}$ is the dielectric constant at high energy, and the middle and the last terms are the Drude and Lorentz components, respectively. The Drude term represents for itinerate electrons while the Lorentz terms are used to describe excitations across energy gaps and interband transitions. The spectrum of 350 K can be well reproduced by one Drude and two Lorentz terms. In contrast, two additional Lorentz terms have to be added to reproduce the two peaks at 3600 \cm and 2100\cm due to the opening of two energy gaps, indicated by dotted lines in Figure 2 (b). Meanwhile, the spectral weight of Drude component is substantially removed. The resonance frequencies of the last two Lorentz components are temperature independent, indicating that they should be ascribed to interband transitions. The fitting parameters of 350 K and 10 K are shown in Table 1.

It is well known that the square of plasma frequency $\omega_p$ is proportional to $n/m^*$, where n is the number of free carriers and $m^*$ is the effective mass of electrons. Drude-lorentz fitting yield $\omega_p$ to be 50000 \cm at 350 K and 23000 \cm at 10 K. Assuming that the effective mass of free carriers is constant at different temperatures, the remaining FS area at 10 K is only 21\% of that at 350 K. Namely, 79\% of the free carriers are removed away because of the two CDW gaps. The width of the Drude peak indicates the scattering rate of the free carriers $\gamma=1/\tau$, where $\tau$ is the average life time of free carriers. It is shown in Table 1 that $\gamma_D \approx$ 1530 \cm at 350 K but drops sharply to $\gamma_D \approx$ 230 \cm at the lowest temperature. Even though the FS area are mostly gapped away, the scattering rate of the residual free carriers decreases in a greater magnitude, which explains why the conductivity is getting even larger upon temperature cooling.

The above conclusion is against with many other experimental results which indicated only one phase CDW order.\cite{PhysRevB.77.235104,PhysRevB.74.125115,PhysRevLett.101.246402} In order to further substantiate the two CDW gaps revealed by optical spectroscopy, we performed ultrafast pump-probe spectroscopy, which has been proven to be a very effective tool in discerning small energy gaps\cite{PhysRevLett.99.147008,PhysRevLett.82.4918,PhysRevLett.104.027003}. A Ti sapphire oscillator was employed, which produces 800 nm laser pulses with 100 fs width and 80 MHz repetition rate. The pump and probe beams were cross polarized, and an additional polarizer was mounted just in front of the detector in order to eliminate the influence of stray light. The pump intensity was set to be $\sim2 \mu J/cm^2$ and the probe intensity was 10 times lower.
\begin{figure}[htbp]
\includegraphics[width=7.5cm]{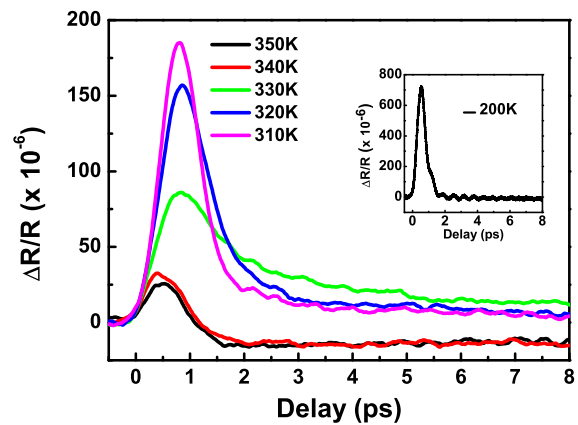}
\caption{(Color online) The photoinduced reflectance of TbTe$_3$ throughout the first phase transition. The inset shows transient reflectance at 200 K. The spectra are slightly smoothed for the sake of better observation.}
\label{Fig:4}
\end{figure}

Figure 3 displays the photoinduced change of reflectance $\Delta R/R$ below and above the first CDW phase transition temperature. Each spectrum consists of a fast initially rise, due to temperature change induced by pump pulses, and a subsequent picoseconds relaxation which can be fit by a single exponential function $\Delta R/R=A exp(-t/\tau)$, where $A$ is the amplitude of transient reflectance and $\tau$ is the relaxation time for photon excited carriers decaying to their original states. The extremely long relaxation dominated by thermal diffusion is not considered here except for the fitting procedure. Above the transition temperature $T_{C1}$, amplitude $A$ and relaxation time $\tau$ are almost temperature independent, while both of them get enlarged in a great magnitude by entering the CDW state. Moreover, below $T_{C1}$ the amplitude increases upon temperature cooling while the relaxation time evolves in an opposite fashion. As will be explained in detail later, this is a powerful proof of energy gap opening. The temperature dependent $A(T)$ and $\tau(T)$ can be extracted by single exponential fitting for various temperatures, which are plotted in figure 4 (a) and (b) respectively. $A(T)$ rises sharply at 331 K and the relaxation time $\tau(T)$ shows a quasidivergence at the same temperature, in associate with the first CDW transition. As temperature decreasing, a sudden upturn emerges at 165 K in the $A(T)$ plot, and a peak like feature develops subsequently. This behavior is not observed in the previous pump-probe measurement\cite{PhysRevLett.101.246402}, but quite resembles to what appears below $T_{C1}$ only that there is no clear corresponding anomaly in the relaxation time $\tau(T)$.

Actually the single exponential fitting of photoinduced reflectance is not so perfect as expected due to the appearance of oscillations by entering the CDW state. In order to show the oscillations more vividly, we plot the spectrum at 200 K, taken under a pump fluence of $\sim 8 \mu J$, in the inset of Figure 3. The frequencies could be obtained through Fourier transformation of the net waves, which are yielded by subtracting the single exponential part of the spectrum. Within the limitation of resolution, we found the 2.1 THz amplitude mode and a 1.8 THz phonon as reported by R.V. Yusupov et al.\cite{PhysRevLett.101.246402}

For the purpose of better understanding, we use Rothwarf-Taylor (RT) model\cite{PhysRevLett.19.27} to interpret the results and try to offer more information. RT model is a phenomenological model initially being proposed to explain the ultrafast relaxation mechanism in superconductors, according to which the SC energy gap would give rise to a bottleneck to the decay of photon excited quasiparticles (QPs). When the pump beam illuminating on the sample, Cooper pairs would be broken and excited over the SC gap to higher energy levels. Thus a large number of QPs are created, the decaying of which would generate phonons with energy $\omega > 2\Delta$. The high frequency phonons could break more Cooper pairs in return, bringing about avalanche to the relaxation of QPs. Therefore a long relaxation time is expected corresponding to the energy gap formation. Based on this model, the density of excited quasiparticles $n_T$ can be gotten via the amplitude of $\Delta R/R$, $n_T\propto[A(T)/A(T\rightarrow0)]^{-1}-1$. Then in associate with the thermal quasipaticle density $n_T\propto\sqrt{\Delta(T)T}exp(-\Delta(T)/T)$, $\Delta(0)$ can be acquired by\cite{Kabanov1999,PhysRevLett.83.800}
\begin{equation}
A(T)\propto \frac{\varepsilon_I/(\Delta(T)+\kappa_BT/2)}{1+\gamma\sqrt{\frac{2\kappa_BT}{\pi\Delta(T)}}exp(-\Delta(T)/\kappa_BT)}
\end{equation}
in which $\varepsilon_I$ is the pump intensity and $\gamma$ is a fitting parameter.

Although RT model is proposed to explain the bottleneck effect in a superconductor due to the opening of pairing energy gap, it is widely used to interpret similar effects of decay time increase in CDW materials (e.g. K0.3MoO3,\cite{PhysRevLett.83.800} RTe3,\cite{PhysRevLett.101.246402} transition metal dichacogenides\cite{Demsar2002}бн) or even in heavy fermion systems with the presence of hybridization energy gaps\cite{Chia2011,Qi2013}. We use equation (2) to fit the temperature dependent amplitude and the result is shown in Figure 4 (a). Since the energy scale of the first CDW gap is extremely large as revealed by optical spectroscopy, phonons or other bosonic excitations with comparable energy are definitely absent. Assuming that $\Delta (T)$ obeys the BCS temperature dependence, RT model is operational only in a narrow temperature range just blow $T_C$. The fit procedure above 200 K yields $\gamma=10$ and $2\Delta_{1}=14\kappa_BT_{C1}$, roughly in consistent with our infrared measurement, applying $\Delta(T)=\Delta(0)\sqrt{1-T/T_C}$.

The most outstanding feature of $\tau(T)$ is a quasidivergence at 331 K. According to RT model, the relaxation rate near $T_C$ is dominated by the energy transfer from high frequency phonons with $\omega > 2\Delta$ to low frequency phonons with $\omega < 2\Delta$, whereas the electron-phonon collisions could be neglected. The phonon relaxation rate can be express as\cite{Kabanov1999}
\begin{equation}
\frac{1}{\tau_{ph}}=\frac{12\Gamma_\omega\kappa_BT^{'}\Delta(T)}{\hbar\omega^2}
\end{equation}
where $\Gamma_\omega$ represents for the Raman phonon linewidth and $T^{'}$ is the QP temperature. Apparently, $\tau^{-1}(T) \propto \Delta (T)$ in the vicinity of $T_C$, thus a divergence is expected at the phase transition temperature where energy gap starts to open. Consequently, the sudden increase in $A(T)$ and quasidivergence in $\tau(T)$ illustrate unambiguously the appearance of an energy gap, although the transition temperature identified here is a little bit lower than its real value because of heating effect. As shown by the red line in Figure 4 (b), $\tau(T)$ is well reproduced by equation (3).
\begin{figure}[htbp]
\includegraphics[width=7.5cm]{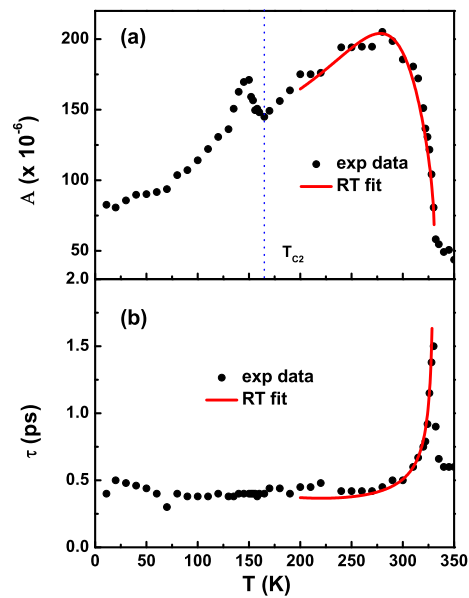}
\caption{(Color online) Temperature dependent (a) amplitude $A(T)$ and (b) relaxation time $\tau(T)$. The solid lines are fitting results according to RT model.}
\label{Fig:4}
\end{figure}

Apart from that, the sudden upturn appearing at 165 K could be ascribed to another gap even though there is no divergence near $T_{C2}$ in the relaxation time plot. Similar behaviors have been disclosed by ultrafast pump-probe spectroscopy for HoTe$_3$ and DyTe$_3$,\cite{PhysRevLett.101.246402} both of which experience two CDW transitions. Since only a small part of the Fermi surface becomes gapped through the second transition, which is reasonable considering its tiny spectral weight in $\sigma_1(\omega)$, the influence of this gap to the relaxation process could be neglected and the supposed divergence at $T_{C2}$ would be too weak to be detected. Another thing needs to be noticed here is that the real phase transition temperature should be a little bit higher than the measured value (165 K), due to heating effect.

Previous report of the X-ray diffraction\cite{PhysRevB.77.035114} in TbTe$_3$ also provides clues on the existence of a second
gap. The temperature dependent order parameter exhibits an apparent dip at approximately 150 K, which was considered as an experimental artifact by the authors. As a contrast, for ErTe$_3$ which is believed to experience two CDW phase transitions, the small dip in order parameter plot is viewed as the interaction between the two anomalies. Based on our experimental observations, we believe that the dip feature of TbTe$_3$ should indicate a second phase transition as well.

Distinct from our results, the ARPES measurements conducted on TbTe$_3$ by Schmitt et al.\cite{Schmitt2011} at 100 K show only the first CDW transition. This may be caused by the limitation of energy and momentum resolution. Compare to the first phase transition which actually removed most part of the low frequency spectral weight in optical conductivity, the new CDW order at lower transition temperature is a much weaker feature. Therefore, ARPES measurements at 100 K might be too high to get good enough energy and momentum resolution. Furthermore, one has to be very careful in searching for the region of energy gap formation. Apparently, further study in ARPES is needed on resolving the issue specifically.

Before conclusion, we shall elaborate the difference between the second CDW order observed here and the one
reported by Banerjee et al. at much lower transition temperature.\cite{PhysRevB.87.155131} As we mentioned, it is well known that the RTe$_3$ system with R ions heavier than Tb (Dy, Ho, Er, Tm) experiences two CDW transitions, both of which originate from the nesting of FS, with wave vectors of almost the same magnitude but perpendicular to each other. Moreover, the second phase transition temperature decreases as the radii of the R ions increasing. The second CDW phase transition
reported by Banerjee et al. at 41 K follows this trend. By contrast, the CDW phase transition near 165 K is completely out of this trend. In an earlier optical spectroscopy study on one of the heavy rare-earth element based compound ErTe$_3$ (with $T_{C1}$=267 K and $T_{C2}$=150 K), we clearly observed the development of the second energy gap at the energy scale of $\sim890$ \cm.\cite{PhysRevB.84.155132} If the second CDW phase transition at 41 K is present in our TbTe$_3$ sample, we should observe the energy gap at even lower energy scale. However, in our experiment, no signature of this phase transition was seen. Instead, we find an energy gap feature at much higher energy scale (2100 \cm or 260 meV) and the identified transition temperature by pump-probe experiment is also much higher, severely violating the revolution trend of heavy RTe$_3$s. Obviously, the newly discovered gap in this work is out of any catalog that is already known. It is worth noting that the infrared measurement of CeTe$_3$\cite{PhysRevB.83.155113} shows similar results as TbTe$_3$, with a second CDW transition emerging below 200 K. These two materials may have the same mechanism, which is mysterious and requires for more systematic and meticulous experiments in future.

\section{conclusion}

To conclude, we have performed optical spectroscopy and ultrafast pump-probe measurements on the single crystalline TbTe$_3$ compound. Optical conductivity demonstrates clearly the formation of two CDW gaps with the energy scales of 3600 \cm (450 meV) and 2100 \cm (260 meV), respectively. The ultrafast dynamics of QPs confirmed the existence of two CDW orders and identified the transition temperature to be $T_{C1}$ =336 K and $T_{C2}$ =165 K. The higher temperature one is well known before while the lower one is never discovered in any other experiments. This new CDW order does not follow the trend of second CDW order observed for the heavy rare-earth element based RTe$_3$ compounds in both transition temperature and energy scale of gap, therefore it should have a different origin from the RTe$_3$ compounds with heavier R ions.

\begin{center}
\small{\textbf{ACKNOWLEDGMENTS}}
\end{center}

This work was supported by the National Science Foundation of
China (11120101003, 11327806, 11074291), and the 973 project of the Ministry of Science and Technology of China (2011CB921701, 2012CB821403).

\bibliographystyle{apsrev4-1}

\begin{thebibliography}{26}%
\makeatletter
\providecommand \@ifxundefined [1]{%
 \@ifx{#1\undefined}
}%
\providecommand \@ifnum [1]{%
 \ifnum #1\expandafter \@firstoftwo
 \else \expandafter \@secondoftwo
 \fi
}%
\providecommand \@ifx [1]{%
 \ifx #1\expandafter \@firstoftwo
 \else \expandafter \@secondoftwo
 \fi
}%
\providecommand \natexlab [1]{#1}%
\providecommand \enquote  [1]{``#1''}%
\providecommand \bibnamefont  [1]{#1}%
\providecommand \bibfnamefont [1]{#1}%
\providecommand \citenamefont [1]{#1}%
\providecommand \href@noop [0]{\@secondoftwo}%
\providecommand \href [0]{\begingroup \@sanitize@url \@href}%
\providecommand \@href[1]{\@@startlink{#1}\@@href}%
\providecommand \@@href[1]{\endgroup#1\@@endlink}%
\providecommand \@sanitize@url [0]{\catcode `\\12\catcode `\$12\catcode
  `\&12\catcode `\#12\catcode `\^12\catcode `\_12\catcode `\%12\relax}%
\providecommand \@@startlink[1]{}%
\providecommand \@@endlink[0]{}%
\providecommand \url  [0]{\begingroup\@sanitize@url \@url }%
\providecommand \@url [1]{\endgroup\@href {#1}{\urlprefix }}%
\providecommand \urlprefix  [0]{URL }%
\providecommand \Eprint [0]{\href }%
\providecommand \doibase [0]{http://dx.doi.org/}%
\providecommand \selectlanguage [0]{\@gobble}%
\providecommand \bibinfo  [0]{\@secondoftwo}%
\providecommand \bibfield  [0]{\@secondoftwo}%
\providecommand \translation [1]{[#1]}%
\providecommand \BibitemOpen [0]{}%
\providecommand \bibitemStop [0]{}%
\providecommand \bibitemNoStop [0]{.\EOS\space}%
\providecommand \EOS [0]{\spacefactor3000\relax}%
\providecommand \BibitemShut  [1]{\csname bibitem#1\endcsname}%
\let\auto@bib@innerbib\@empty
\bibitem [{\citenamefont {Travaglini}\ and\ \citenamefont
  {Wachter}(1984)}]{PhysRevB.30.1971}%
  \BibitemOpen
  \bibfield  {author} {\bibinfo {author} {\bibfnamefont {G.}~\bibnamefont
  {Travaglini}}\ and\ \bibinfo {author} {\bibfnamefont {P.}~\bibnamefont
  {Wachter}},\ }\href {\doibase 10.1103/PhysRevB.30.1971} {\bibfield  {journal}
  {\bibinfo  {journal} {Phys. Rev. B}\ }\textbf {\bibinfo {volume} {30}},\
  \bibinfo {pages} {1971} (\bibinfo {year} {1984})}\BibitemShut {NoStop}%
\bibitem [{\citenamefont {WU}\ and\ \citenamefont {LIEBER}(1989)}]{WU31031989}%
  \BibitemOpen
  \bibfield  {author} {\bibinfo {author} {\bibfnamefont {X.~L.}\ \bibnamefont
  {WU}}\ and\ \bibinfo {author} {\bibfnamefont {C.~M.}\ \bibnamefont
  {LIEBER}},\ }\href {\doibase 10.1126/science.243.4899.1703} {\bibfield
  {journal} {\bibinfo  {journal} {Science}\ }\textbf {\bibinfo {volume}
  {243}},\ \bibinfo {pages} {1703} (\bibinfo {year} {1989})}\BibitemShut
  {NoStop}%
\bibitem [{\citenamefont {Norling}\ and\ \citenamefont
  {Steinfink}(1966)}]{doi:10.1021/ic50043a004}%
  \BibitemOpen
  \bibfield  {author} {\bibinfo {author} {\bibfnamefont {B.~K.}\ \bibnamefont
  {Norling}}\ and\ \bibinfo {author} {\bibfnamefont {H.}~\bibnamefont
  {Steinfink}},\ }\href {\doibase 10.1021/ic50043a004} {\bibfield  {journal}
  {\bibinfo  {journal} {Inorganic Chemistry}\ }\textbf {\bibinfo {volume}
  {5}},\ \bibinfo {pages} {1488} (\bibinfo {year} {1966})}\BibitemShut
  {NoStop}%
\bibitem [{\citenamefont {Laverock}\ \emph {et~al.}(2005)\citenamefont
  {Laverock}, \citenamefont {Dugdale}, \citenamefont {Major}, \citenamefont
  {Alam}, \citenamefont {Ru}, \citenamefont {Fisher}, \citenamefont {Santi},\
  and\ \citenamefont {Bruno}}]{PhysRevB.71.085114}%
  \BibitemOpen
  \bibfield  {author} {\bibinfo {author} {\bibfnamefont {J.}~\bibnamefont
  {Laverock}}, \bibinfo {author} {\bibfnamefont {S.~B.}\ \bibnamefont
  {Dugdale}}, \bibinfo {author} {\bibfnamefont {Z.}~\bibnamefont {Major}},
  \bibinfo {author} {\bibfnamefont {M.~A.}\ \bibnamefont {Alam}}, \bibinfo
  {author} {\bibfnamefont {N.}~\bibnamefont {Ru}}, \bibinfo {author}
  {\bibfnamefont {I.~R.}\ \bibnamefont {Fisher}}, \bibinfo {author}
  {\bibfnamefont {G.}~\bibnamefont {Santi}}, \ and\ \bibinfo {author}
  {\bibfnamefont {E.}~\bibnamefont {Bruno}},\ }\href {\doibase
  10.1103/PhysRevB.71.085114} {\bibfield  {journal} {\bibinfo  {journal} {Phys.
  Rev. B}\ }\textbf {\bibinfo {volume} {71}},\ \bibinfo {pages} {085114}
  (\bibinfo {year} {2005})}\BibitemShut {NoStop}%
\bibitem [{\citenamefont {DiMasi}\ \emph {et~al.}(1995)\citenamefont {DiMasi},
  \citenamefont {Aronson}, \citenamefont {Mansfield}, \citenamefont {Foran},\
  and\ \citenamefont {Lee}}]{PhysRevB.52.14516}%
  \BibitemOpen
  \bibfield  {author} {\bibinfo {author} {\bibfnamefont {E.}~\bibnamefont
  {DiMasi}}, \bibinfo {author} {\bibfnamefont {M.~C.}\ \bibnamefont {Aronson}},
  \bibinfo {author} {\bibfnamefont {J.~F.}\ \bibnamefont {Mansfield}}, \bibinfo
  {author} {\bibfnamefont {B.}~\bibnamefont {Foran}}, \ and\ \bibinfo {author}
  {\bibfnamefont {S.}~\bibnamefont {Lee}},\ }\href {\doibase
  10.1103/PhysRevB.52.14516} {\bibfield  {journal} {\bibinfo  {journal} {Phys.
  Rev. B}\ }\textbf {\bibinfo {volume} {52}},\ \bibinfo {pages} {14516}
  (\bibinfo {year} {1995})}\BibitemShut {NoStop}%
\bibitem [{\citenamefont {DiMasi}\ \emph {et~al.}(1994)\citenamefont {DiMasi},
  \citenamefont {Foran}, \citenamefont {Aronson},\ and\ \citenamefont
  {Lee}}]{doi:10.1021/cm00046a049}%
  \BibitemOpen
  \bibfield  {author} {\bibinfo {author} {\bibfnamefont {E.}~\bibnamefont
  {DiMasi}}, \bibinfo {author} {\bibfnamefont {B.}~\bibnamefont {Foran}},
  \bibinfo {author} {\bibfnamefont {M.~C.}\ \bibnamefont {Aronson}}, \ and\
  \bibinfo {author} {\bibfnamefont {S.}~\bibnamefont {Lee}},\ }\href {\doibase
  10.1021/cm00046a049} {\bibfield  {journal} {\bibinfo  {journal} {Chemistry of
  Materials}\ }\textbf {\bibinfo {volume} {6}},\ \bibinfo {pages} {1867}
  (\bibinfo {year} {1994})}\BibitemShut {NoStop}%
\bibitem [{\citenamefont {Ru}\ \emph {et~al.}(2008)\citenamefont {Ru},
  \citenamefont {Condron}, \citenamefont {Margulis}, \citenamefont {Shin},
  \citenamefont {Laverock}, \citenamefont {Dugdale}, \citenamefont {Toney},\
  and\ \citenamefont {Fisher}}]{PhysRevB.77.035114}%
  \BibitemOpen
  \bibfield  {author} {\bibinfo {author} {\bibfnamefont {N.}~\bibnamefont
  {Ru}}, \bibinfo {author} {\bibfnamefont {C.~L.}\ \bibnamefont {Condron}},
  \bibinfo {author} {\bibfnamefont {G.~Y.}\ \bibnamefont {Margulis}}, \bibinfo
  {author} {\bibfnamefont {K.~Y.}\ \bibnamefont {Shin}}, \bibinfo {author}
  {\bibfnamefont {J.}~\bibnamefont {Laverock}}, \bibinfo {author}
  {\bibfnamefont {S.~B.}\ \bibnamefont {Dugdale}}, \bibinfo {author}
  {\bibfnamefont {M.~F.}\ \bibnamefont {Toney}}, \ and\ \bibinfo {author}
  {\bibfnamefont {I.~R.}\ \bibnamefont {Fisher}},\ }\href {\doibase
  10.1103/PhysRevB.77.035114} {\bibfield  {journal} {\bibinfo  {journal} {Phys.
  Rev. B}\ }\textbf {\bibinfo {volume} {77}},\ \bibinfo {pages} {035114}
  (\bibinfo {year} {2008})}\BibitemShut {NoStop}%
\bibitem [{\citenamefont {Tomic}\ \emph {et~al.}(2009)\citenamefont {Tomic},
  \citenamefont {Rak}, \citenamefont {Veazey}, \citenamefont {Malliakas},
  \citenamefont {Mahanti}, \citenamefont {Kanatzidis},\ and\ \citenamefont
  {Tessmer}}]{PhysRevB.79.085422}%
  \BibitemOpen
  \bibfield  {author} {\bibinfo {author} {\bibfnamefont {A.}~\bibnamefont
  {Tomic}}, \bibinfo {author} {\bibfnamefont {Z.}~\bibnamefont {Rak}}, \bibinfo
  {author} {\bibfnamefont {J.~P.}\ \bibnamefont {Veazey}}, \bibinfo {author}
  {\bibfnamefont {C.~D.}\ \bibnamefont {Malliakas}}, \bibinfo {author}
  {\bibfnamefont {S.~D.}\ \bibnamefont {Mahanti}}, \bibinfo {author}
  {\bibfnamefont {M.~G.}\ \bibnamefont {Kanatzidis}}, \ and\ \bibinfo {author}
  {\bibfnamefont {S.~H.}\ \bibnamefont {Tessmer}},\ }\href {\doibase
  10.1103/PhysRevB.79.085422} {\bibfield  {journal} {\bibinfo  {journal} {Phys.
  Rev. B}\ }\textbf {\bibinfo {volume} {79}},\ \bibinfo {pages} {085422}
  (\bibinfo {year} {2009})}\BibitemShut {NoStop}%
\bibitem [{\citenamefont {Brouet}\ \emph {et~al.}(2008)\citenamefont {Brouet},
  \citenamefont {Yang}, \citenamefont {Zhou}, \citenamefont {Hussain},
  \citenamefont {Moore}, \citenamefont {He}, \citenamefont {Lu}, \citenamefont
  {Shen}, \citenamefont {Laverock}, \citenamefont {Dugdale}, \citenamefont
  {Ru},\ and\ \citenamefont {Fisher}}]{PhysRevB.77.235104}%
  \BibitemOpen
  \bibfield  {author} {\bibinfo {author} {\bibfnamefont {V.}~\bibnamefont
  {Brouet}}, \bibinfo {author} {\bibfnamefont {W.~L.}\ \bibnamefont {Yang}},
  \bibinfo {author} {\bibfnamefont {X.~J.}\ \bibnamefont {Zhou}}, \bibinfo
  {author} {\bibfnamefont {Z.}~\bibnamefont {Hussain}}, \bibinfo {author}
  {\bibfnamefont {R.~G.}\ \bibnamefont {Moore}}, \bibinfo {author}
  {\bibfnamefont {R.}~\bibnamefont {He}}, \bibinfo {author} {\bibfnamefont
  {D.~H.}\ \bibnamefont {Lu}}, \bibinfo {author} {\bibfnamefont {Z.~X.}\
  \bibnamefont {Shen}}, \bibinfo {author} {\bibfnamefont {J.}~\bibnamefont
  {Laverock}}, \bibinfo {author} {\bibfnamefont {S.~B.}\ \bibnamefont
  {Dugdale}}, \bibinfo {author} {\bibfnamefont {N.}~\bibnamefont {Ru}}, \ and\
  \bibinfo {author} {\bibfnamefont {I.~R.}\ \bibnamefont {Fisher}},\ }\href
  {\doibase 10.1103/PhysRevB.77.235104} {\bibfield  {journal} {\bibinfo
  {journal} {Phys. Rev. B}\ }\textbf {\bibinfo {volume} {77}},\ \bibinfo
  {pages} {235104} (\bibinfo {year} {2008})}\BibitemShut {NoStop}%
\bibitem [{\citenamefont {Gweon}\ \emph {et~al.}(1998)\citenamefont {Gweon},
  \citenamefont {Denlinger}, \citenamefont {Clack}, \citenamefont {Allen},
  \citenamefont {Olson}, \citenamefont {DiMasi}, \citenamefont {Aronson},
  \citenamefont {Foran},\ and\ \citenamefont {Lee}}]{PhysRevLett.81.886}%
  \BibitemOpen
  \bibfield  {author} {\bibinfo {author} {\bibfnamefont {G.-H.}\ \bibnamefont
  {Gweon}}, \bibinfo {author} {\bibfnamefont {J.~D.}\ \bibnamefont
  {Denlinger}}, \bibinfo {author} {\bibfnamefont {J.~A.}\ \bibnamefont
  {Clack}}, \bibinfo {author} {\bibfnamefont {J.~W.}\ \bibnamefont {Allen}},
  \bibinfo {author} {\bibfnamefont {C.~G.}\ \bibnamefont {Olson}}, \bibinfo
  {author} {\bibfnamefont {E.}~\bibnamefont {DiMasi}}, \bibinfo {author}
  {\bibfnamefont {M.~C.}\ \bibnamefont {Aronson}}, \bibinfo {author}
  {\bibfnamefont {B.}~\bibnamefont {Foran}}, \ and\ \bibinfo {author}
  {\bibfnamefont {S.}~\bibnamefont {Lee}},\ }\href {\doibase
  10.1103/PhysRevLett.81.886} {\bibfield  {journal} {\bibinfo  {journal} {Phys.
  Rev. Lett.}\ }\textbf {\bibinfo {volume} {81}},\ \bibinfo {pages} {886}
  (\bibinfo {year} {1998})}\BibitemShut {NoStop}%
\bibitem [{\citenamefont {Moore}\ \emph {et~al.}(2010)\citenamefont {Moore},
  \citenamefont {Brouet}, \citenamefont {He}, \citenamefont {Lu}, \citenamefont
  {Ru}, \citenamefont {Chu}, \citenamefont {Fisher},\ and\ \citenamefont
  {Shen}}]{PhysRevB.81.073102}%
  \BibitemOpen
  \bibfield  {author} {\bibinfo {author} {\bibfnamefont {R.~G.}\ \bibnamefont
  {Moore}}, \bibinfo {author} {\bibfnamefont {V.}~\bibnamefont {Brouet}},
  \bibinfo {author} {\bibfnamefont {R.}~\bibnamefont {He}}, \bibinfo {author}
  {\bibfnamefont {D.~H.}\ \bibnamefont {Lu}}, \bibinfo {author} {\bibfnamefont
  {N.}~\bibnamefont {Ru}}, \bibinfo {author} {\bibfnamefont {J.-H.}\
  \bibnamefont {Chu}}, \bibinfo {author} {\bibfnamefont {I.~R.}\ \bibnamefont
  {Fisher}}, \ and\ \bibinfo {author} {\bibfnamefont {Z.-X.}\ \bibnamefont
  {Shen}},\ }\href {\doibase 10.1103/PhysRevB.81.073102} {\bibfield  {journal}
  {\bibinfo  {journal} {Phys. Rev. B}\ }\textbf {\bibinfo {volume} {81}},\
  \bibinfo {pages} {073102} (\bibinfo {year} {2010})}\BibitemShut {NoStop}%
\bibitem [{\citenamefont {Sacchetti}\ \emph {et~al.}(2006)\citenamefont
  {Sacchetti}, \citenamefont {Degiorgi}, \citenamefont {Giamarchi},
  \citenamefont {Ru},\ and\ \citenamefont {Fisher}}]{PhysRevB.74.125115}%
  \BibitemOpen
  \bibfield  {author} {\bibinfo {author} {\bibfnamefont {A.}~\bibnamefont
  {Sacchetti}}, \bibinfo {author} {\bibfnamefont {L.}~\bibnamefont {Degiorgi}},
  \bibinfo {author} {\bibfnamefont {T.}~\bibnamefont {Giamarchi}}, \bibinfo
  {author} {\bibfnamefont {N.}~\bibnamefont {Ru}}, \ and\ \bibinfo {author}
  {\bibfnamefont {I.~R.}\ \bibnamefont {Fisher}},\ }\href {\doibase
  10.1103/PhysRevB.74.125115} {\bibfield  {journal} {\bibinfo  {journal} {Phys.
  Rev. B}\ }\textbf {\bibinfo {volume} {74}},\ \bibinfo {pages} {125115}
  (\bibinfo {year} {2006})}\BibitemShut {NoStop}%
\bibitem [{\citenamefont {Yusupov}\ \emph {et~al.}(2008)\citenamefont
  {Yusupov}, \citenamefont {Mertelj}, \citenamefont {Chu}, \citenamefont
  {Fisher},\ and\ \citenamefont {Mihailovic}}]{PhysRevLett.101.246402}%
  \BibitemOpen
  \bibfield  {author} {\bibinfo {author} {\bibfnamefont {R.~V.}\ \bibnamefont
  {Yusupov}}, \bibinfo {author} {\bibfnamefont {T.}~\bibnamefont {Mertelj}},
  \bibinfo {author} {\bibfnamefont {J.-H.}\ \bibnamefont {Chu}}, \bibinfo
  {author} {\bibfnamefont {I.~R.}\ \bibnamefont {Fisher}}, \ and\ \bibinfo
  {author} {\bibfnamefont {D.}~\bibnamefont {Mihailovic}},\ }\href {\doibase
  10.1103/PhysRevLett.101.246402} {\bibfield  {journal} {\bibinfo  {journal}
  {Phys. Rev. Lett.}\ }\textbf {\bibinfo {volume} {101}},\ \bibinfo {pages}
  {246402} (\bibinfo {year} {2008})}\BibitemShut {NoStop}%
\bibitem [{\citenamefont {Banerjee}\ \emph {et~al.}(2013)\citenamefont
  {Banerjee}, \citenamefont {Feng}, \citenamefont {Silevitch}, \citenamefont
  {Wang}, \citenamefont {Lang}, \citenamefont {Kuo}, \citenamefont {Fisher},\
  and\ \citenamefont {Rosenbaum}}]{PhysRevB.87.155131}%
  \BibitemOpen
  \bibfield  {author} {\bibinfo {author} {\bibfnamefont {A.}~\bibnamefont
  {Banerjee}}, \bibinfo {author} {\bibfnamefont {Y.}~\bibnamefont {Feng}},
  \bibinfo {author} {\bibfnamefont {D.~M.}\ \bibnamefont {Silevitch}}, \bibinfo
  {author} {\bibfnamefont {J.}~\bibnamefont {Wang}}, \bibinfo {author}
  {\bibfnamefont {J.~C.}\ \bibnamefont {Lang}}, \bibinfo {author}
  {\bibfnamefont {H.-H.}\ \bibnamefont {Kuo}}, \bibinfo {author} {\bibfnamefont
  {I.~R.}\ \bibnamefont {Fisher}}, \ and\ \bibinfo {author} {\bibfnamefont
  {T.~F.}\ \bibnamefont {Rosenbaum}},\ }\href {\doibase
  10.1103/PhysRevB.87.155131} {\bibfield  {journal} {\bibinfo  {journal} {Phys.
  Rev. B}\ }\textbf {\bibinfo {volume} {87}},\ \bibinfo {pages} {155131}
  (\bibinfo {year} {2013})}\BibitemShut {NoStop}%
\bibitem [{\citenamefont {Chia}\ \emph {et~al.}(2007)\citenamefont {Chia},
  \citenamefont {Zhu}, \citenamefont {Talbayev}, \citenamefont {Averitt},
  \citenamefont {Taylor}, \citenamefont {Oh}, \citenamefont {Jo},\ and\
  \citenamefont {Lee}}]{PhysRevLett.99.147008}%
  \BibitemOpen
  \bibfield  {author} {\bibinfo {author} {\bibfnamefont {E.~E.~M.}\
  \bibnamefont {Chia}}, \bibinfo {author} {\bibfnamefont {J.-X.}\ \bibnamefont
  {Zhu}}, \bibinfo {author} {\bibfnamefont {D.}~\bibnamefont {Talbayev}},
  \bibinfo {author} {\bibfnamefont {R.~D.}\ \bibnamefont {Averitt}}, \bibinfo
  {author} {\bibfnamefont {A.~J.}\ \bibnamefont {Taylor}}, \bibinfo {author}
  {\bibfnamefont {K.-H.}\ \bibnamefont {Oh}}, \bibinfo {author} {\bibfnamefont
  {I.-S.}\ \bibnamefont {Jo}}, \ and\ \bibinfo {author} {\bibfnamefont {S.-I.}\
  \bibnamefont {Lee}},\ }\href {\doibase 10.1103/PhysRevLett.99.147008}
  {\bibfield  {journal} {\bibinfo  {journal} {Phys. Rev. Lett.}\ }\textbf
  {\bibinfo {volume} {99}},\ \bibinfo {pages} {147008} (\bibinfo {year}
  {2007})}\BibitemShut {NoStop}%
\bibitem [{\citenamefont {Demsar}\ \emph
  {et~al.}(1999{\natexlab{a}})\citenamefont {Demsar}, \citenamefont {Podobnik},
  \citenamefont {Kabanov}, \citenamefont {Wolf},\ and\ \citenamefont
  {Mihailovic}}]{PhysRevLett.82.4918}%
  \BibitemOpen
  \bibfield  {author} {\bibinfo {author} {\bibfnamefont {J.}~\bibnamefont
  {Demsar}}, \bibinfo {author} {\bibfnamefont {B.}~\bibnamefont {Podobnik}},
  \bibinfo {author} {\bibfnamefont {V.~V.}\ \bibnamefont {Kabanov}}, \bibinfo
  {author} {\bibfnamefont {T.}~\bibnamefont {Wolf}}, \ and\ \bibinfo {author}
  {\bibfnamefont {D.}~\bibnamefont {Mihailovic}},\ }\href {\doibase
  10.1103/PhysRevLett.82.4918} {\bibfield  {journal} {\bibinfo  {journal}
  {Phys. Rev. Lett.}\ }\textbf {\bibinfo {volume} {82}},\ \bibinfo {pages}
  {4918} (\bibinfo {year} {1999}{\natexlab{a}})}\BibitemShut {NoStop}%
\bibitem [{\citenamefont {Chia}\ \emph {et~al.}(2010)\citenamefont {Chia},
  \citenamefont {Talbayev}, \citenamefont {Zhu}, \citenamefont {Yuan},
  \citenamefont {Park}, \citenamefont {Thompson}, \citenamefont {Panagopoulos},
  \citenamefont {Chen}, \citenamefont {Luo}, \citenamefont {Wang},\ and\
  \citenamefont {Taylor}}]{PhysRevLett.104.027003}%
  \BibitemOpen
  \bibfield  {author} {\bibinfo {author} {\bibfnamefont {E.~E.~M.}\
  \bibnamefont {Chia}}, \bibinfo {author} {\bibfnamefont {D.}~\bibnamefont
  {Talbayev}}, \bibinfo {author} {\bibfnamefont {J.-X.}\ \bibnamefont {Zhu}},
  \bibinfo {author} {\bibfnamefont {H.~Q.}\ \bibnamefont {Yuan}}, \bibinfo
  {author} {\bibfnamefont {T.}~\bibnamefont {Park}}, \bibinfo {author}
  {\bibfnamefont {J.~D.}\ \bibnamefont {Thompson}}, \bibinfo {author}
  {\bibfnamefont {C.}~\bibnamefont {Panagopoulos}}, \bibinfo {author}
  {\bibfnamefont {G.~F.}\ \bibnamefont {Chen}}, \bibinfo {author}
  {\bibfnamefont {J.~L.}\ \bibnamefont {Luo}}, \bibinfo {author} {\bibfnamefont
  {N.~L.}\ \bibnamefont {Wang}}, \ and\ \bibinfo {author} {\bibfnamefont
  {A.~J.}\ \bibnamefont {Taylor}},\ }\href {\doibase
  10.1103/PhysRevLett.104.027003} {\bibfield  {journal} {\bibinfo  {journal}
  {Phys. Rev. Lett.}\ }\textbf {\bibinfo {volume} {104}},\ \bibinfo {pages}
  {027003} (\bibinfo {year} {2010})}\BibitemShut {NoStop}%
\bibitem [{\citenamefont {Rothwarf}\ and\ \citenamefont
  {Taylor}(1967)}]{PhysRevLett.19.27}%
  \BibitemOpen
  \bibfield  {author} {\bibinfo {author} {\bibfnamefont {A.}~\bibnamefont
  {Rothwarf}}\ and\ \bibinfo {author} {\bibfnamefont {B.~N.}\ \bibnamefont
  {Taylor}},\ }\href {\doibase 10.1103/PhysRevLett.19.27} {\bibfield  {journal}
  {\bibinfo  {journal} {Phys. Rev. Lett.}\ }\textbf {\bibinfo {volume} {19}},\
  \bibinfo {pages} {27} (\bibinfo {year} {1967})}\BibitemShut {NoStop}%
\bibitem [{\citenamefont {Kabanov}\ \emph {et~al.}(1999)\citenamefont
  {Kabanov}, \citenamefont {Demsar}, \citenamefont {Podobnik},\ and\
  \citenamefont {Mihailovic}}]{Kabanov1999}%
  \BibitemOpen
  \bibfield  {author} {\bibinfo {author} {\bibfnamefont {V.~V.}\ \bibnamefont
  {Kabanov}}, \bibinfo {author} {\bibfnamefont {J.}~\bibnamefont {Demsar}},
  \bibinfo {author} {\bibfnamefont {B.}~\bibnamefont {Podobnik}}, \ and\
  \bibinfo {author} {\bibfnamefont {D.}~\bibnamefont {Mihailovic}},\ }\href
  {\doibase 10.1103/PhysRevB.59.1497} {\bibfield  {journal} {\bibinfo
  {journal} {Phys. Rev. B}\ }\textbf {\bibinfo {volume} {59}},\ \bibinfo
  {pages} {1497} (\bibinfo {year} {1999})}\BibitemShut {NoStop}%
\bibitem [{\citenamefont {Demsar}\ \emph
  {et~al.}(1999{\natexlab{b}})\citenamefont {Demsar}, \citenamefont
  {Biljakovi\ifmmode~\acute{c}\else \'{c}\fi{}},\ and\ \citenamefont
  {Mihailovic}}]{PhysRevLett.83.800}%
  \BibitemOpen
  \bibfield  {author} {\bibinfo {author} {\bibfnamefont {J.}~\bibnamefont
  {Demsar}}, \bibinfo {author} {\bibfnamefont {K.}~\bibnamefont
  {Biljakovi\ifmmode~\acute{c}\else \'{c}\fi{}}}, \ and\ \bibinfo {author}
  {\bibfnamefont {D.}~\bibnamefont {Mihailovic}},\ }\href {\doibase
  10.1103/PhysRevLett.83.800} {\bibfield  {journal} {\bibinfo  {journal} {Phys.
  Rev. Lett.}\ }\textbf {\bibinfo {volume} {83}},\ \bibinfo {pages} {800}
  (\bibinfo {year} {1999}{\natexlab{b}})}\BibitemShut {NoStop}%
\bibitem [{\citenamefont {Demsar}\ \emph {et~al.}(2002)\citenamefont {Demsar},
  \citenamefont {Forr\'o}, \citenamefont {Berger},\ and\ \citenamefont
  {Mihailovic}}]{Demsar2002}%
  \BibitemOpen
  \bibfield  {author} {\bibinfo {author} {\bibfnamefont {J.}~\bibnamefont
  {Demsar}}, \bibinfo {author} {\bibfnamefont {L.}~\bibnamefont {Forr\'o}},
  \bibinfo {author} {\bibfnamefont {H.}~\bibnamefont {Berger}}, \ and\ \bibinfo
  {author} {\bibfnamefont {D.}~\bibnamefont {Mihailovic}},\ }\href {\doibase
  10.1103/PhysRevB.66.041101} {\bibfield  {journal} {\bibinfo  {journal} {Phys.
  Rev. B}\ }\textbf {\bibinfo {volume} {66}},\ \bibinfo {pages} {041101}
  (\bibinfo {year} {2002})}\BibitemShut {NoStop}%
\bibitem [{\citenamefont {Chia}\ \emph {et~al.}(2011)\citenamefont {Chia},
  \citenamefont {Zhu}, \citenamefont {Talbayev}, \citenamefont {Lee},
  \citenamefont {Hur}, \citenamefont {Moreno}, \citenamefont {Averitt},
  \citenamefont {Sarrao},\ and\ \citenamefont {Taylor}}]{Chia2011}%
  \BibitemOpen
  \bibfield  {author} {\bibinfo {author} {\bibfnamefont {E.~E.~M.}\
  \bibnamefont {Chia}}, \bibinfo {author} {\bibfnamefont {J.-X.}\ \bibnamefont
  {Zhu}}, \bibinfo {author} {\bibfnamefont {D.}~\bibnamefont {Talbayev}},
  \bibinfo {author} {\bibfnamefont {H.~J.}\ \bibnamefont {Lee}}, \bibinfo
  {author} {\bibfnamefont {N.}~\bibnamefont {Hur}}, \bibinfo {author}
  {\bibfnamefont {N.~O.}\ \bibnamefont {Moreno}}, \bibinfo {author}
  {\bibfnamefont {R.~D.}\ \bibnamefont {Averitt}}, \bibinfo {author}
  {\bibfnamefont {J.~L.}\ \bibnamefont {Sarrao}}, \ and\ \bibinfo {author}
  {\bibfnamefont {A.~J.}\ \bibnamefont {Taylor}},\ }\href {\doibase
  10.1103/PhysRevB.84.174412} {\bibfield  {journal} {\bibinfo  {journal} {Phys.
  Rev. B}\ }\textbf {\bibinfo {volume} {84}},\ \bibinfo {pages} {174412}
  (\bibinfo {year} {2011})}\BibitemShut {NoStop}%
\bibitem [{\citenamefont {Qi}\ \emph {et~al.}(2013)\citenamefont {Qi},
  \citenamefont {Durakiewicz}, \citenamefont {Trugman}, \citenamefont {Zhu},
  \citenamefont {Riseborough}, \citenamefont {Baumbach}, \citenamefont {Bauer},
  \citenamefont {Gofryk}, \citenamefont {Meng}, \citenamefont {Joyce},
  \citenamefont {Taylor},\ and\ \citenamefont {Prasankumar}}]{Qi2013}%
  \BibitemOpen
  \bibfield  {author} {\bibinfo {author} {\bibfnamefont {J.}~\bibnamefont
  {Qi}}, \bibinfo {author} {\bibfnamefont {T.}~\bibnamefont {Durakiewicz}},
  \bibinfo {author} {\bibfnamefont {S.~A.}\ \bibnamefont {Trugman}}, \bibinfo
  {author} {\bibfnamefont {J.-X.}\ \bibnamefont {Zhu}}, \bibinfo {author}
  {\bibfnamefont {P.~S.}\ \bibnamefont {Riseborough}}, \bibinfo {author}
  {\bibfnamefont {R.}~\bibnamefont {Baumbach}}, \bibinfo {author}
  {\bibfnamefont {E.~D.}\ \bibnamefont {Bauer}}, \bibinfo {author}
  {\bibfnamefont {K.}~\bibnamefont {Gofryk}}, \bibinfo {author} {\bibfnamefont
  {J.-Q.}\ \bibnamefont {Meng}}, \bibinfo {author} {\bibfnamefont {J.~J.}\
  \bibnamefont {Joyce}}, \bibinfo {author} {\bibfnamefont {A.~J.}\ \bibnamefont
  {Taylor}}, \ and\ \bibinfo {author} {\bibfnamefont {R.~P.}\ \bibnamefont
  {Prasankumar}},\ }\href {\doibase 10.1103/PhysRevLett.111.057402} {\bibfield
  {journal} {\bibinfo  {journal} {Phys. Rev. Lett.}\ }\textbf {\bibinfo
  {volume} {111}},\ \bibinfo {pages} {057402} (\bibinfo {year}
  {2013})}\BibitemShut {NoStop}%
\bibitem [{\citenamefont {Schmitt}\ \emph {et~al.}(2011)\citenamefont
  {Schmitt}, \citenamefont {Kirchmann}, \citenamefont {Bovensiepen},
  \citenamefont {Moore}, \citenamefont {Chu}, \citenamefont {Lu}, \citenamefont
  {Rettig}, \citenamefont {Wolf}, \citenamefont {Fisher},\ and\ \citenamefont
  {Shen}}]{Schmitt2011}%
  \BibitemOpen
  \bibfield  {author} {\bibinfo {author} {\bibfnamefont {F.}~\bibnamefont
  {Schmitt}}, \bibinfo {author} {\bibfnamefont {P.~S.}\ \bibnamefont
  {Kirchmann}}, \bibinfo {author} {\bibfnamefont {U.}~\bibnamefont
  {Bovensiepen}}, \bibinfo {author} {\bibfnamefont {R.~G.}\ \bibnamefont
  {Moore}}, \bibinfo {author} {\bibfnamefont {J.-H.}\ \bibnamefont {Chu}},
  \bibinfo {author} {\bibfnamefont {D.~H.}\ \bibnamefont {Lu}}, \bibinfo
  {author} {\bibfnamefont {L.}~\bibnamefont {Rettig}}, \bibinfo {author}
  {\bibfnamefont {M.}~\bibnamefont {Wolf}}, \bibinfo {author} {\bibfnamefont
  {I.~R.}\ \bibnamefont {Fisher}}, \ and\ \bibinfo {author} {\bibfnamefont
  {Z.-X.}\ \bibnamefont {Shen}},\ }\href {\doibase
  10.1088/1367-2630/13/6/063022} {\bibfield  {journal} {\bibinfo  {journal}
  {New Journal of Physics}\ }\textbf {\bibinfo {volume} {13}},\ \bibinfo
  {pages} {063022} (\bibinfo {year} {2011})}\BibitemShut {NoStop}%
\bibitem [{\citenamefont {Hu}\ \emph {et~al.}(2011{\natexlab{a}})\citenamefont
  {Hu}, \citenamefont {Cheng}, \citenamefont {Yuan}, \citenamefont {Dong},
  \citenamefont {Fang}, \citenamefont {Guo}, \citenamefont {Chen},
  \citenamefont {Zheng}, \citenamefont {Shi},\ and\ \citenamefont
  {Wang}}]{PhysRevB.84.155132}%
  \BibitemOpen
  \bibfield  {author} {\bibinfo {author} {\bibfnamefont {B.~F.}\ \bibnamefont
  {Hu}}, \bibinfo {author} {\bibfnamefont {B.}~\bibnamefont {Cheng}}, \bibinfo
  {author} {\bibfnamefont {R.~H.}\ \bibnamefont {Yuan}}, \bibinfo {author}
  {\bibfnamefont {T.}~\bibnamefont {Dong}}, \bibinfo {author} {\bibfnamefont
  {A.~F.}\ \bibnamefont {Fang}}, \bibinfo {author} {\bibfnamefont {W.~T.}\
  \bibnamefont {Guo}}, \bibinfo {author} {\bibfnamefont {Z.~G.}\ \bibnamefont
  {Chen}}, \bibinfo {author} {\bibfnamefont {P.}~\bibnamefont {Zheng}},
  \bibinfo {author} {\bibfnamefont {Y.~G.}\ \bibnamefont {Shi}}, \ and\
  \bibinfo {author} {\bibfnamefont {N.~L.}\ \bibnamefont {Wang}},\ }\href
  {\doibase 10.1103/PhysRevB.84.155132} {\bibfield  {journal} {\bibinfo
  {journal} {Phys. Rev. B}\ }\textbf {\bibinfo {volume} {84}},\ \bibinfo
  {pages} {155132} (\bibinfo {year} {2011}{\natexlab{a}})}\BibitemShut
  {NoStop}%
\bibitem [{\citenamefont {Hu}\ \emph {et~al.}(2011{\natexlab{b}})\citenamefont
  {Hu}, \citenamefont {Zheng}, \citenamefont {Yuan}, \citenamefont {Dong},
  \citenamefont {Cheng}, \citenamefont {Chen},\ and\ \citenamefont
  {Wang}}]{PhysRevB.83.155113}%
  \BibitemOpen
  \bibfield  {author} {\bibinfo {author} {\bibfnamefont {B.~F.}\ \bibnamefont
  {Hu}}, \bibinfo {author} {\bibfnamefont {P.}~\bibnamefont {Zheng}}, \bibinfo
  {author} {\bibfnamefont {R.~H.}\ \bibnamefont {Yuan}}, \bibinfo {author}
  {\bibfnamefont {T.}~\bibnamefont {Dong}}, \bibinfo {author} {\bibfnamefont
  {B.}~\bibnamefont {Cheng}}, \bibinfo {author} {\bibfnamefont {Z.~G.}\
  \bibnamefont {Chen}}, \ and\ \bibinfo {author} {\bibfnamefont {N.~L.}\
  \bibnamefont {Wang}},\ }\href {\doibase 10.1103/PhysRevB.83.155113}
  {\bibfield  {journal} {\bibinfo  {journal} {Phys. Rev. B}\ }\textbf {\bibinfo
  {volume} {83}},\ \bibinfo {pages} {155113} (\bibinfo {year}
  {2011}{\natexlab{b}})}\BibitemShut {NoStop}%
\end{thebibliography}

\end{document}